\documentclass[12pt]{iopart}
\usepackage{graphicx}
\expandafter\let\csname equation*\endcsname\relax
\expandafter\let\csname endequation*\endcsname\relax
\usepackage{amsmath}
\usepackage{mathabx}
\usepackage{cite}
\usepackage{lscape}
\usepackage{varwidth}
\usepackage{textcomp}
\usepackage{xcolor,soul}
\usepackage{color,soul}
\usepackage{algorithm}
\usepackage{algpseudocode}
\usepackage{bm}
\usepackage{esvect}
\usepackage{accents}
\usepackage{caption}
\usepackage[symbol]{footmisc}
\usepackage{lmodern}

\begin{document}

\title[MICROSCOPE mission scenario]{MICROSCOPE Mission scenario, ground segment and data processing}

\author{Manuel Rodrigues$^1$, Pierre Touboul$^1$ \footnote[2]{\,Deceased in February 2021}, Gilles M\'etris$^2$, Judicael Bedouet$^3$, Joel Berg\'e$^1$, Patrice Carle$^4$, Ratana Chhun$^1$, Bruno Christophe$^1$, Bernard Foulon$^1$, Pierre-Yves Guidotti$^4$ \footnote{\,Current address: AIRBUS Defence and Space, F-31402 Toulouse, France}, Stephanie Lala$^5$, Alain Robert$^4$}

\address{$^1$ DPHY, ONERA, Universit\'e Paris Saclay, F-92322 Ch\^atillon, France}
\address{$^2$ Universit\'e C\^ote d{'}Azur, Observatoire de la C\^ote d'Azur, CNRS, IRD, G\'eoazur, 250 avenue Albert Einstein, F-06560 Valbonne, France}
\address{$^3$ ONERA Midi-Pyr\'en\'ees, 18 avenue E. Belin, F-31400 Toulouse, France}
\address{$^4$ CNES, 18 avenue E. Belin, F-31400 Toulouse, France}
\address{$^5$ ONERA, Universit\'e Paris Saclay, F-91123 Palaiseau, France}

\ead{manuel.rodrigues@onera.fr, gilles.metris@oca.eu}
\vspace{10pt}
\begin{indented}
\item[] Aug. 2021
\end{indented}

\begin{abstract}
Testing the Weak Equivalence Principle (WEP) to a precision of $10^{-15}$ requires a quantity of data that give enough confidence on the final result: ideally, the longer the measurement the better the rejection of thestatistical noise. The science sessions had a duration of 120 orbits maximum and were regularly repeated and spaced out to accommodate operational constraints but also in order to repeat the experiment in different conditions and to allow time to calibrate the instrument. Several science sessions were performed over the 2.5 year duration of the experiment. This paper aims to describe how the data have been produced on the basis of a mission scenario and a data flow process, driven by a tradeoff between the science objectives and the operational constraints.
The mission was led by the Centre National d'Etudes Spatiales (CNES) which provided the satellite, the launch and the ground operations. The ground segment was distributed between CNES and Office National d'Etudes et de Recherches A\'erospatiales (ONERA). CNES provided the raw data through the Centre d'Expertise de Compensation de Tra\^{i}n\'{e}e (CECT: Drag-free expertise centre). The science was led by the Observatoire de la C\^ote d{'}Azur (OCA) and ONERA was in charge of the data process. The latter also provided the instrument and the Science Mission Centre of MICROSCOPE (CMSM).

\end{abstract}


\submitto{\CQG}
%
%
%

\section{Introduction}

Theories aiming to extend the standard model or to modify General Relativity (e.g. to unify both realms) are difficult to test \cite{will14}. However, most of them predict a violation of the WEP, providing an experimental window into them. At the end of the 20th century, new technologies, such as drag-free satellites \cite{triad1} and ultra sensitive accelerometers, allowed for a leap forward in the precision of tests of the WEP (or, equivalently, of the universality of free-fall). First concepts for such tests were developped in 1971 \cite{chapman01} and followed by the STEP mission aiming to test the EP to $10^{-18}$ thanks to cryogenic temperature operations \cite{everitt2003, sumner2007, Overduin12}. Taking advantage of the CNES MYRIAD program opportunity, MICROSCOPE (Micro-Satellite \`a tra\^in\'ee Compens\'ee pour l'Observation du Principe d'Equivalence) is a space mission operating at room temperature and built on those new technologies, dedicated to the test of the Weak Equivalence Principle (WEP) with an accuracy objective of $10^{-15}$. Refs. \cite{touboul17, touboul19} gave the first result on the basis of 120 orbits (8.3 days). This result has been improved and widely discussed in Ref. \cite{metriscqg9}.Testing the WEP consist in comparing  the normalised difference of acceleration of two bodies in the same gravity field \cite{rodriguescqg1}:
\begin{equation} \label{eq_eotvos}
\eta (2,1) = 2 \frac{a_2-a_1}{a_2+a_1} = 2 \frac{m_{g2}/m_{i2} - m_{g1}/m_{i1}}{m_{g2}/m_{i2} + m_{g1}/m_{i1}}
\end{equation}
where $a_j$ is the acceleration of the $j$th test-body, and $m_{g,j}$ and $m_{i,j}$ are its gravitational and inertial masses. Usually, in  MICROSCOPE papers, because $m_g/m_i$ is already constrained to unity at $10^{-13}$ level, a good first order approximation of the E\"otv\"os parameter, $\eta (2,1)$, is taken as $\delta (2,1)$ when comparing material 1 and 2:
\begin{equation}
\delta(2,1) \equiv \frac{m_{g2}}{m_{i2}} - \frac{m_{g1}}{m_{i1}}= \eta (1+\mathcal{O}(10^{-13})).
\end{equation}
 
The MICROSCOPE platform is a micro-satellite inherited from the CNES MYRIAD satellite program \cite{robertcqg3}. The T-SAGE (Twin Space Accelerometer for Gravitation Experiment) is the unique payload and scientific instrument on board. It was developed by ONERA and integrated in the satellite's core by CNES. It comprised two sensor units (SU) (Refs. \cite{liorzoucqg2, robertcqg3, rodriguescqg1} give detailed descriptions of the instrument, of the satellite and of the measurement equation). Each sensor unit is a double concentric accelerometer. The first one (called SUEP for its role as the sensor unit measuring the WEP) is composed of two concentric test-masses made of different materials: platinum-rhodium alloy with 90\% of Pt and 10\% of Rh in mass for one material and titanium alloy with 90\% of Ti, 6\% of Al and 4\% of V  for the other material \cite{touboul19}. The second sensor (called SUREF for its role as a reference sensor unit) contains two concentric test-masses made of the same material (PtRh10) and is expected to give a null signal.  SUREF helps to check the overall data production and processes. Associated to each SU, a front-end electronic unit (FEEU) is integrated with the SU in the satellite thermal and magnetic shielded cocoon. The FEEU provides all the precise and stable reference voltages necessary to the accelerometers. It also delivers the voltage applied to the electrodes surrounding the test-masses in order to control their motion. An interface control unit (ICU) integrated on one of the wall of the satellite receives the capacitive position sensor information from the FEEU and computes the control electrode voltages to be transmitted to the FEEU. It also delivers to the satellite all the science and housekeeping data of the instrument for the ground telemetry. 

\begin{figure*}
\includegraphics[width=0.45 \textwidth]{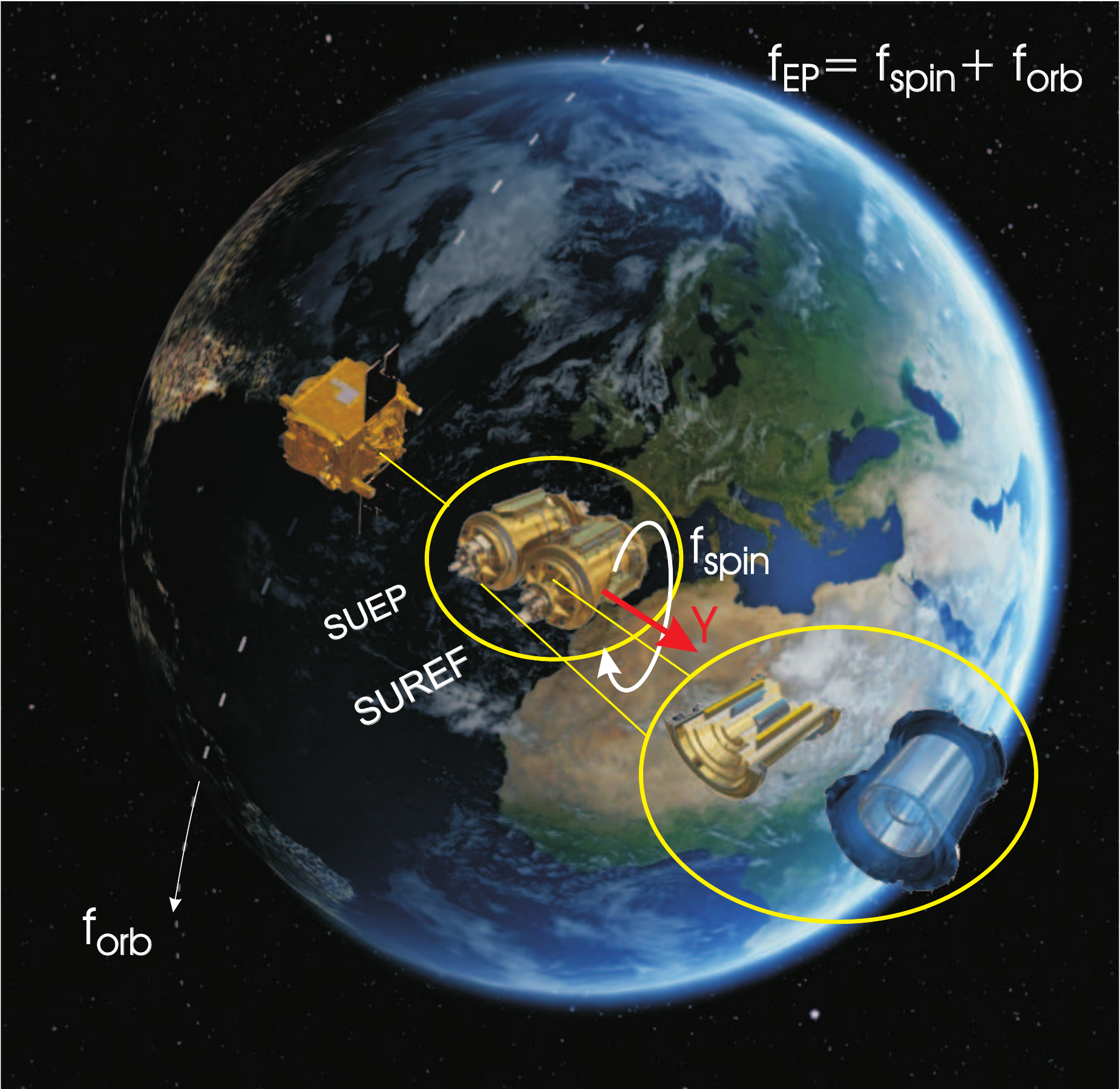}
\includegraphics[width=0.453 \textwidth]{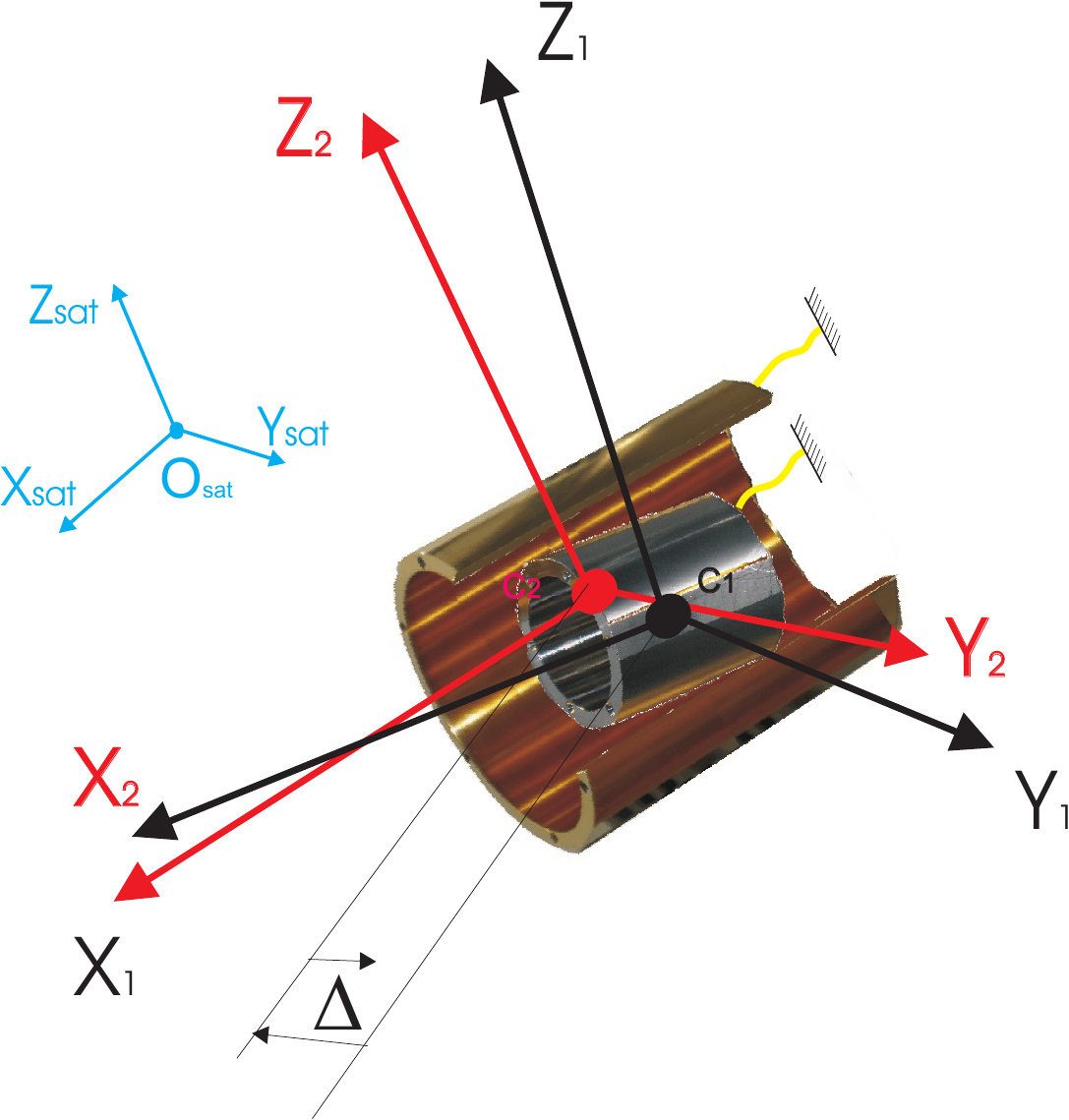}
\caption{Left: the  4 test-masses orbiting around the Earth (credits CNES / Virtual-IT 2017) . Right: test-masses and satellite frames; the ($X_{\rm sat}$, $Y_{\rm sat}$, $Z_{\rm sat}$) triad defines the satellite frame; the reference frames ($X_k$, $Y_k$, $Z_k$, $k=1,2$) are attached to the test-masses (black for the inner mass $k=1$, red for the outer mass $k=2$).}
\label{fig_orbit} 
\end{figure*}

Finally, the accelerometers are also used on board, in addition to the star-trackers data, to drive the Drag-Free and Attitude Control System (DFACS) \cite{robertcqg3}. The DFACS compensates the external disturbing forces and torques in order to minimise the common mode accelerations \cite{rodriguescqg1}. 
The instrument's reference frame is defined as follows (see \Fref{fig_orbit}): the $X$ and $Z$ axes are maintained in the orbital plane and the $Y$ axis is normal to it, in all configurations of the EP test. The main (most sensitive) measurement axis is the $X$ axis.

The satellite and its payload can operate in different configurations (e.g. inertial or rotating pointing of the satellite, full range or high resolution of the instrument, under various temperature changes,...) which have been tested and validated on ground. A session is a period during which the configuration is kept unchanged. The mission scenario is defined as the succession of different sessions.

In this paper, comprehensive information is given to detail the way the data are produced. From the ground segment organisation described in \Sref{sec_ground}, an overview of the drivers that led to built the science mission scenario is presented in \Sref{missdriv}. The realised scenario is then described in \Sref{scen}. On the basis of this scenario, the data pipeline is described in \Sref{sect_data}.
The WEP analysis algorithm is discussed in Refs. \cite{bergecqg8, metriscqg9}.


\section{Ground segment} \label{sec_ground}

\begin{figure}
\includegraphics[width=0.9 \textwidth]{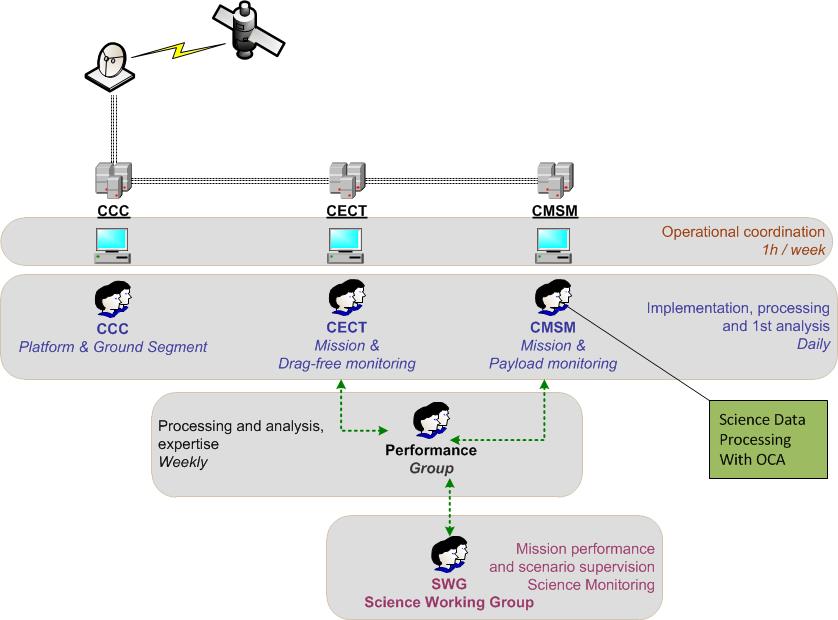}
\caption{Schematic of the ground segment organisation.}
\label{fig_ground}       
\end{figure}

The ground segment (\Fref{fig_ground}) was distributed over CNES's and ONERA's sites in France. CNES led the operations in Toulouse at the Control Command Centre (CCC) by communicating with the satellite through ground stations spread over the Earth. The Centre d'Expertise de Compensation de Tra\^{i}n\'{e}e (CECT), also located at CNES Toulouse, was in charge of interfacing ONERA's Science Mission Center (CMSM -- Centre de Mission Science de MICROSCOPE) and CCC. The CMSM was developed and operated in close collaboration between the two institutes ONERA and OCA.

Based on science objectives, CMSM defined a weekly mission scenario and submitted it to CECT for validation. In collaboration with CCC, CECT then prepared the program files to be sent to the satellite.
The CECT managed the operational constraints and all on-board activities (satellite maintenance, survival modes, eclipse and Moon periods, technological experiments). Beyond this operational function, the CECT expert team provided a daily monitoring of the DFACS and evaluated the satellite performance every two weeks \cite{robertcqg3}. In addition to the CECT monitoring, CMSM monitored the payload daily and estimated the mission performance in order to validate the executed scientific sessions. 

The performance group (composed of CNES, OCA, ONERA and ZARM members) analysed monthly the status of the mission and of the data processing related to the performance of the mission. 

Finally, the status of the mission was presented in a yearly basis to the Science Working Group committee (SWG) composed of scientists specialising in areas related to the MICROSCOPE mission, who did not take part in the day-to-day operations of the mission. The SWG advised the MICROSCOPE operational and science team on the best strategy for the mission scenario. It also reviewed the science data process during the mission and after the end of the satellite operation until the publication of the final results.


\section{Mission scenario drivers} \label{missdriv}
\subsection {Nominal drivers}

\subsubsection{Eclipses and Moon phases}

The mission was designed for a duration of one year with one year of extension. The selected sun-synchronous orbit allowed for a very stable thermal configuration, except during three months of eclipses about the summer solstice. During this period of poorer thermal stability, it was decided to perform only satellite maintenance operations or experiments that did not require high accuracy.
After the launch, at the end of April 2016, the commissioning phase was performed during the first following eclipse period. 

Beside eclipses, ``full Moon'' periods prevented running science measurement. The satellite had to be re-pointed during four out of every 28 days to avoid blinding of the star tracker sensors by the Moon light.  Consequently, calibration sessions requiring satellite oscillations were performed far from these full Moon periods in order to keep margins on the acceptable angle of oscillation. This re-pointing also generated a thermal variation that reduced the performance of the instrument and thus the relevance of science measurements performed during these periods.  

Each EP session had as long duration as possible to minimise the statistical error, with a maximum duration set to 120 orbits by the operational constraints of the DFACS. To maximise the time spent in science mode, the operational scenario was built to limit the number of transitions between different types of sessions. Indeed, different types of session mean different control laws for the DFACS and thus additional time to make these control algorithms converge. Two orbits have been allocated to these transitions to allow the controller to converge. Transition data are not used in the science data process.
Finally, each kind of session was designed and verified on ground on an end-to-end real time test bench \cite{bergecqg7}, each session type independently of the others.

\subsubsection{Gas consumption}

The gas consumption was also an obvious driver of the mission scenario. The satellite contained about 16~kg of nitrogen, mainly used by the DFACS in the science sessions. The main consumption contributor comes from the V3 satellite configuration in which the satellite rotates about the axis normal to the orbital plane with the highest rotation frequency $f_{\rm{spin_3}} = 2.94315 \times{}10^{-3}~{\rm Hz}$. The V2 configuration corresponds to a lower rate rotation frequency defined by  $f_{\rm{spin_2}} = 0.75681 \times{}10^{-3}~{\rm Hz}$. \Tref{tab_Gas} gives the typical consumption measured in gram per orbit for the different satellite configurations. Some margins were taken into account for emergency manoeuvres (e.g. collision avoidance manoeuvres). Other margins were also considered for unexpected events that could make the satellite enter into survival mode: this actually happened 12 times over the two years of operation, mostly related to Single Event Upsets (SEU) in the DFACS subcomponents (star trackers, accelerometers, gas thruster electronics). Calibration sessions were short (typically 5-orbit-long) and performed in quasi-inertial pointing leading to little impact on the gas consumption. 

\begin{table}
\caption{\label{tab_Gas} Cold gas consumption measured in orbit}
\begin{indented}
\item[]\begin{tabular}{@{}lrr}
\br
Satellite configuration & mean consumption & total consumption\\
 & in g/orbit & in g\\
\mr
Inertial (excluding calibrations) & 1.2 & 2 300\\
Calibration in quasi-inertial pointing & 5.0 & 1 400\\ 
Rotation in V2 mode & 1.2 & 1 100\\
Rotation in V3 mode & 6.6 &  10 700\\
Transitions & 1.2 & 700\\
\br
\end{tabular}
\end{indented}
\end{table}

\subsubsection{Mission scenario}

With all these constraints taken into account, the mission scenario was built on the basis of elementary sessions executed successively. This succession of sessions was programmed on board for a period of 15 days and updated weekly. The process of management of the scenario was part of the ground segment operation. Thanks to the robust operation of the drag-free system, very few sessions had to be re-programmed.

\subsection {Impact of housekeeping capacitance failures on the scenario}

The mission operation had to deal with three capacitance failures in the SUREF. The first one occurred in May 2016, three weeks after the launch, the second failure in June 2016 and the third one in February 2018. The first and third failures occurred on a voltage housekeeping capacitance. The second one was linked to a capacitance failure in the generation of the 10~ kHz reference voltage (see Ref. \cite{liorzoucqg2} for details on the role of this voltage). Fortunately, all these failures had no impact on the performance of the capacitive sensor in flight. This was confirmed by ground tests with engineering models and numerical simulations. However, each failure increased the power consumption by 2~W in SUREF's FEEU, thereby inducing a temperature increase. 

It was found that a series of flight capacitances had been weakened during the circuit board production. All these capacitances were potentially likely to fail in SUREF or SUEP electronics. After the second failure, in order to minimise risks, the mission scenario was constrained as follows:

\begin{itemize}
\item turn on only one SU at a time to minimise the temperature of operation as the risk increased with temperature. Operating the two SU units, i.e. the four test-masses, simultaneously would have been ideal for cross-checking but not mandatory for the science measurements because only one SU is in an optimal drag-free environment controlled by the DFACS; 
\item switch off both SU when acceleration measurements are not needed for science, as the risk increased with time of operation: the lifetime of each SU had to be preserved; 
\item assign priority to performing EP science tests with the SUEP only and postpone all other minor in-flight operations. Fortunately the SUEP proved to be very robust.
\end{itemize}

These constraints allowed for the successful operation of both SU during the entire mission. In February 2018, after completion of most of initially planned science scenario, the two SU were switched on simultaneously and provided good results during several days. However, the third failure occurred and confirmed the hypothesis of the risk increasing with a higher temperature (the operation of both SU at a time makes the temperature reach 35\textdegree{}C instead of less than 10\textdegree{}C with only one). The simultaneous switch-on of both SU was eventually discontinued.


\section{Implemented scenario} \label{scen}

\begin{figure}
\includegraphics[width=0.4 \textwidth]{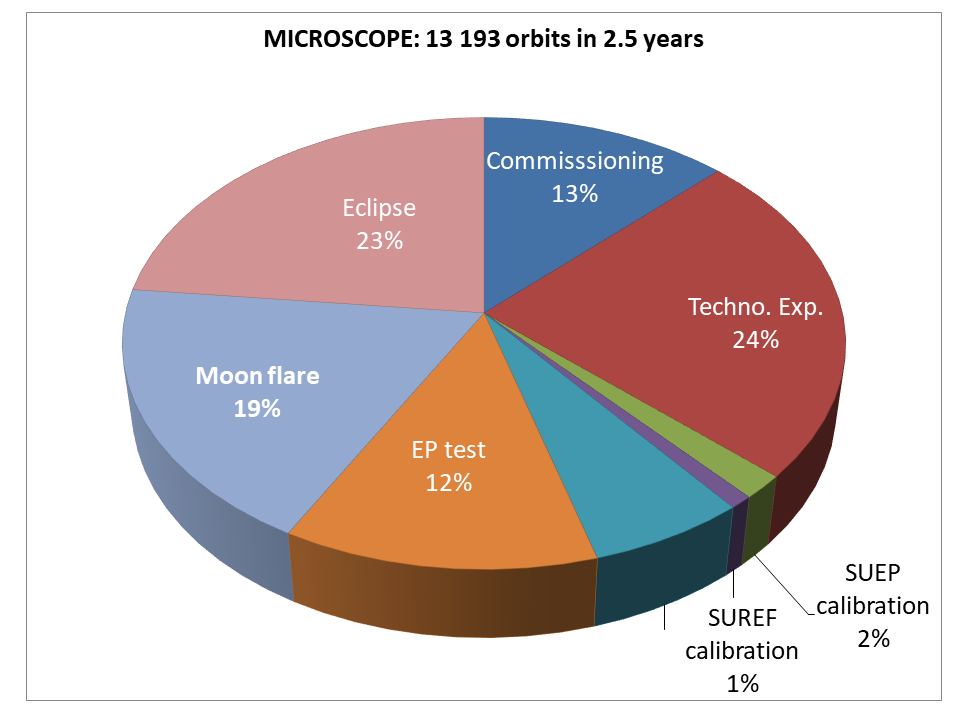}
\includegraphics[width=0.5 \textwidth]{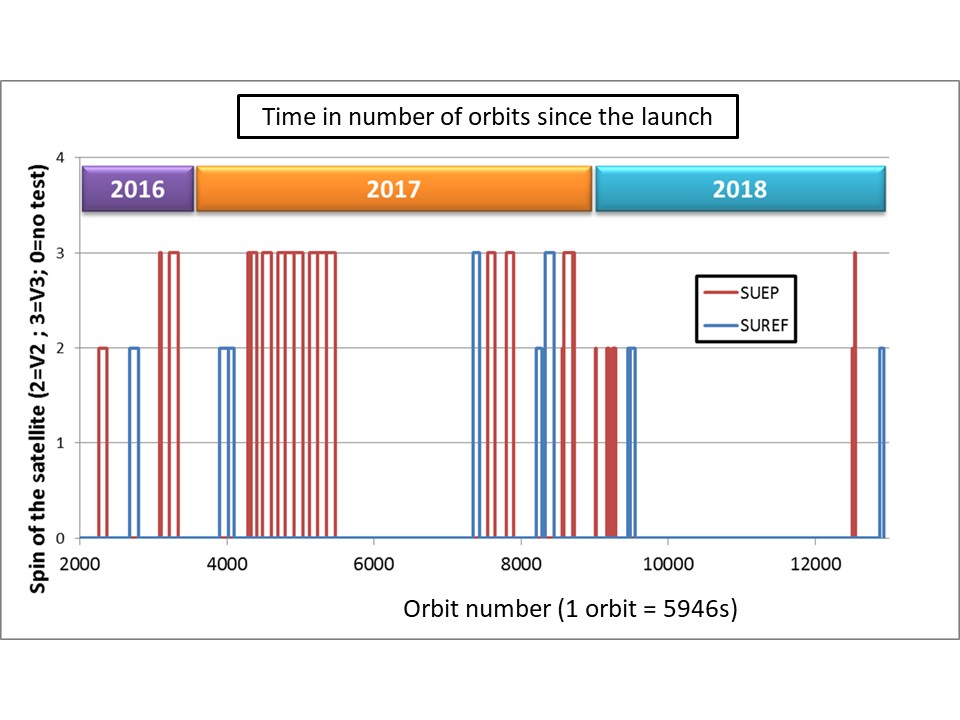}
\centering
\caption{Summary of the realised scenario.}
\label{fig_scen}
\end{figure}

The MICROSCOPE mission lasted 2.5 years, allowing for 13193 orbits (with an orbital period of $5945.9\pm{}0.1~{\rm s}$). The orbits are numbered from the launch date.
Figure \ref{fig_scen} shows the time allocations between different satellite operations during the entire mission. The chart on the left panel provides a general view, while the right panel focuses on the distribution of science sessions throughout the mission, with time defined as the number of orbits since launch.

After the commissioning phase from April to Novembre 2016, science sessions were run until May 2017 (orbits 3000 to 5500, see Figure \ref{fig_scen}). The instrument was switched off between May and August 2017 during the eclipse period, and the second part of the science program was performed from September 2017 to February 2018 (orbits 7300 to 9500). It was followed by a long period of technological sessions during which SUEP was continuously measuring, the risk of failure being less stringent with respect to the gain in instrument characterisation. Finally, short science sessions were performed for noise evaluation in September 2018 nearly at the end of the mission.

A compromise has to be done between performing several consecutive 120-orbit scientific sessions to minimise stochastic errors and performing the necessary satellite maintenance and verifications as well as payload characterisation and calibration. Maintenance and instrument's characterisation sessions were part of the technical sessions.  In addition, the calibration of the instrument detailed in Ref.\cite{hardycqg6} (scale factor, alignment, non-linearities, thermal sensitivities) as well as the evaluation of its stability over time were regularly performed in order to optimise the science output of the mission. These calibration sessions were regularly performed before or after EP test sessions. 

As described in \Sref{missdriv}, the scenario was organised with several sets of EP tests with one SU bracketed by two ``full Moon''. Regularly  encompassed by a scale factor matching (${\rm CAL}_{\rm{K1_{dx}}}$ in \Tref{tab_scii}), one or several EP tests were performed in rotating mode (V2 or V3 -- see below) completed by other calibration sessions (${\rm CAL}_{\rm{delta_{Y}}}$, ${\rm CAL}_{\rm{teta_{dZ}}}$ and ${\rm CAL}_{\rm{teta_{dY}}}$). Very few non-linearity-calibration dedicated sessions were performed because the differential non-linearity could also be deduced from ${\rm CAL}_{\rm{K1_{dx}}}$. The duration of the EP tests was adjusted to the allowed time between two Moon periods, including time for calibration. Allowing for the time necessary to process data and upload a new program, sessions could be replayed if necessary in the following month (after the following Moon period). 

\begin{table}
\caption{\label{tab_scii} Science sessions in quasi-inertial pointing (instrument frame). Sine accelerations have an amplitude of $5\times{}10^{-8}\,m\,s^{-2}$ at $f_{\rm cal}=1.22848 \times{}10^{-3}$ Hz. Satellite oscillations are performed with 0.05\,rad amplitude at $f_{\rm cal}$. Sine modulation are performed by adding a sine signal in the loop of 0.15625\,V at 60\,Hz multiplied by a square signal at 1\,mHz. Detailed description is in Ref. \cite{chhuncqg5}}
\begin{indented}
\item[]\begin{tabular}{@{}llll}
\br
Session & Configuration & Duration & Purpose\cr
\mr
EP$_{\rm{I}}$ & & 17 orbits & EP test (commissioning phase)\cr
 &  & & first performance evaluation\cr
\mr
${\rm CAL}_{\rm{K1_{dx}}}$ & A sine acceleration & 5.07 orbits & Calibration of scale factor \cr
 &  is applied along X & & matching along X\cr
 \mr
${\rm CAL}_{\rm{delta_Y}}$ & Satellite oscillates & 5.07 orbits & Calibration of test-mass relative \cr
 & about  Z  & & offcentring along Y\cr
 \mr
${\rm CAL}_{\rm{teta_{dZ}}}$ & A sine acceleration  & 5.07 orbits & Calibration of misalignment\cr
 & is applied along Y & &between the 2 test-masses\cr
 & & & in the plane X-Z\cr
 \mr
${\rm CAL}_{\rm{teta_{dY}}}$ & A sine acceleration & 5.07 orbits & Calibration of misalignment\cr
 &  is applied along Z  & &between the 2 test-masses\cr
 &  & & in the plane X-Y\cr
\mr
${\rm CAL}_{\rm{K21_{xx}}}$ & Sine signal modulated& 10 orbits & Calibration of non-linear term \cr
  & by a square signal & &  along X\cr
 & on IS1 along X & & \cr
\mr
${\rm CAL}_{\rm{K22_{xx}}}$ & Sine signal modulated& 10 orbits & Calibration of non-linear term \cr
  & by a square signal & & along X\cr
 & on IS2 along X & & \cr
\mr
${\rm PLC}_{\rm{K}}$ & Test mass position & 10 orbits & Calibration of non-linear term \cr
  &biased by a sine signal & & along each axis\cr
\br
\end{tabular}
\end{indented}
\end{table}

The realised mission scenario, succession of several technical and science sessions, is detailed in \ref{sect_tabscen}. 

\subsection{Science sessions} \label{sect:sci}

Science sessions were led by CMSM (all others sessions were led by CECT).
They were performed with several satellite configurations (\Tref{tab_scii} and \Tref{tab_scir}):
\begin{itemize}
\item quasi-inertial pointing: in this mode very near to an inertial pointing, the orbital plane is adjusted slowly by one rotation per year to maintain the orientation of the solar panels towards the sun. The EP test frequency in this mode is simply the orbital frequency $f_{\rm{orb}} = 1.6818 \times{}10^{-4}~{\rm Hz}$; this mode was used mainly during calibration sessions, though some EP tests were performed in ${\rm EP_I}$ during the commissioning phase, but were discarded from the final data analysis \cite{metriscqg9};
\item spinning mode: the satellite rotates about the axis normal to the orbital plane, in the opposite direction to the orbital motion in order to increase the frequency of the EP signal. \Tref{tab_scir} shows the different frequencies of rotation $f_{\rm{spin}}$.  ${\rm EPR_{V1}}$ was used only in commissioning phase. The EP test frequency becomes $f_{\rm{EP}}=f_{\rm{orb}}+f_{\rm{spin}}$. This configuration was the baseline for the EP test.
\end{itemize}

\begin{table} [H]
\caption{\label{tab_scir} Science sessions with satellite rotating about $Y$ axis}
\begin{indented}
\item[]\begin{tabular}{@{}llll}
\br
Session & Satellite spin rate frequency & Duration & Purpose\cr
\mr
$\rm{EPR_{V1}}$ & $f_{\rm{spin_1}}=\frac{7}{2} f_{\rm{orb}}=0.58863 \times{}10^{-3}$ Hz& 20 orbits & commissioning \cr
\mr
$\rm{EPR_{V2}}$ & $f_{\rm{spin_2}}=\frac{9}{2} f_{\rm{orb}}=0.75681 \times{}10^{-3}$ Hz& 20 to & EP test in V2 mode\cr
&&120 orbits&\cr
\mr
$\rm{EPR_{V3}}$ & $f_{\rm{spin_3}}=\frac{35}{2} f_{\rm{orb}}=2.94315 \times{}10^{-3}$ Hz& 20 to & EP test in V3 mode\cr
&&120 orbits&\cr
\br
\end{tabular}
\end{indented}
\end{table}
The evaluation of the E\"otv\"os parameter was performed with sessions lasting longer than 20 orbits in order to sufficiently minimise the statistical noise. 
The satellite rotation frequency $f_{\rm{spin}}$ was selected on two criteria: (i) minimise the spectral leakage of the EP spectral component due to the finite duration of the measurement $T_d$ (i.e. frequency path $f_d=1/T_d$ in the spectral domain) and (ii) avoid the projection of orbital periodicity perturbations at $f_{\rm EP}=1/T_{\rm EP}$. Therefore, we chose $f_{\rm{spin}}=\frac{k}{2}f_{\rm{orb}}$, where $k$ is an odd integer, and $T_d=2 n T_{\rm EP}$ where n is an integer. With this choice, criteria (i) and (ii) are achieved: we have the frequencies $f_{\rm{orb}}$, $f_{\rm{spin}}$, $f_{\rm{EP}}$ and their harmonics multiple of $f_d$ but not multiple each other, and $f_{\rm EP}=f_{\rm spin}+ f_{\rm{orb}}$ is a non-multiple of $f_{\rm{orb}}$.   

To perform a calibration (i.e. estimate an instrumental parameter), a $1.2285\times{}10^{-3}~{\rm Hz}$ ($\frac{73}{10} f_{\rm{orb}}$) sine signal was added on the DFACS loop to bias the linear acceleration or the attitude control, such that the effect of the estimated parameter was increased and concentrated at a well defined frequency. \Tref{tab_scii} lists the different calibration sessions. 
Calibration sessions were less demanding in performance and could afford to be shorter. Initially, the duration of the calibration sessions was fixed to $10/f_{\rm{orb}}$ (73 calibration periods) but  given the very good signal to noise ratio observed in flight and the absence of significant leakage, 5.07 orbits (37 calibration periods) happened to be sufficient.

\Fref{fig_scen} and \Tref{nbsci} summarise the allocated time for each type of session. 1642 orbits were dedicated to the EP test with SUEP and 821 orbits to the SUREF. Until the end of the commissioning phase, the V2 configuration was considered the baseline of EP sessions. However, because of a noise higher than expected (especially for SUEP), the new V3 configuration was defined as the baseline instead of the V2 one by increasing the satellite rotation rate. Thus, most of the tests with SUEP were performed in the V3 configuration (\Tref{tab_scir}), while most tests with SUREF used the V2 configuration. 
V3 configuration resulted in a higher value of $f_{\rm{EP}}$ at which the accelerometer had better performance. At the beginning of the mission, the SUEP noise was evaluated to be less than $11\times{}10^{-11}~{\rm m\,s^{-2}Hz^{-1/2}}$ and SUREF noise to less than $4\times{}10^{-11}~{\rm m\,s^{-2}Hz^{-1/2}}$ at $f_{\rm{EP}}=0.92499 \times{}10^{-3}~{\rm Hz}$ in V2 mode. Although the gas consumption increased by a factor of five in V3, a factor of three could be gained in noise. Thus, nine times less amount of data was required to reach the same statistical error on the science measurement. The trade-off between the mission duration, the gas consumption and the performance led to select the V3 configuration as the baseline for SUEP; as SUREF did not require much improvement, the V2 mode was kept as its baseline. 

\begin{table} [H]
\caption{\label{nbsci} Number of orbits in different satellite configurations for the EP test. One orbit is about 5946\,s. Inertial pointing sessions are not considered}
\begin{indented}
\item[]\begin{tabular}{@{}lcc}
\br
Configuration & \multicolumn{2}{l}{Number of cumulated orbits} \\
 & SUREF & SUEP \\
\mr
$\rm{EPR_{V2}}$ & 563 & 240 \\
\mr
$\rm{EPR_{V3}}$ & 214 & 1402  \\
\br
\end{tabular}
\end{indented}
\end{table}

As shown in Ref. \cite{chhuncqg5}, the noise seems to decrease with time during the whole mission and converge to a stable value. No clear reason was found to this evolution as some parameters (PID essentially) of the instrument or of the satellite (V3 mode) were also changed during the decay period. Therefore, the gain on the ratio performance/gas provided by the V3 configuration compared to the V2 one at the beginning of the mission became less important at the end of the mission. The mean noise over the whole mission was evaluated to be:
\begin{itemize}
\item $(4.0\pm{}0.4)\times{}10^{-11}$\,m\,s$^{-2}$\,Hz$^{-1/2}$ at $3.11133 \times{}10^{-3}$ Hz in V3 mode for SUEP;
\item $(1.8\pm{}1.5)\times{}10^{-11}$\,m\,s$^{-2}$\,Hz$^{-1/2}$ at $0.92499 \times{}10^{-3}$ Hz in V2 mode for SUREF.
\end{itemize}

Taking into account these mean noises and the realised number of orbits, the overall expected statistical error contribution on the E\"otv\"os parameter should be about $1.2\times{}10^{-15}$ for SUREF and $1.7\times{}10^{-15}$ for SUEP. These ballpark numbers are not too far from the final results detailed in Ref. \cite{metriscqg9} as it could be expected for a Gaussian noise. 


\subsection{Technical sessions} \label{sect:tech}

These sessions were led by CECT. Four types of technical sessions were realised during the commissioning phase or during periods not usable for science (\Sref{missdriv}): 
\begin{itemize}
\item instrument characterisation: thermal sensitivity, change of test-mass control law, test-mass free motion range, transfer function, stiffness, coupling tests \cite{chhuncqg5};
\item the satellite characterisation: propulsion, inertia wheel rotation impact on accelerometer, DFACS, GPS, \dots;
\item aeronomy experiment: in 2018, after completion of the science, the SUEP was maintained switched on from February to September 2018 (orbits 9500 to 12500), including the eclipse period, and used for aeronomy measurements alternatively with instrument or satellite characterisation; 
\item satellite maintenance: upload of onboard software updates, reset due to anomalies, test-manoeuvre to avoid a potential collision, propulsion gas tank monitoring, standby, deorbitation at the end of the mission.
\end{itemize}

During technological tests, the DFACS could have three main configurations: off, 3-angular-axis control (fine pointing) or 6-axis control (the science mode being one of the submodes). 


\section {Data production} \label{sect_data}

The data analysis process is detailed in Refs. \cite{bergecqg7, metriscqg9}. The analysis relies on the following information and data: linear and angular acceleration measurements from the accelerometers, satellite position, attitude of the satellite in the inertial reference frame J2000, start time of the session (and by extension all information on the session as the stimuli signals in the DFACS loop, the drag-free test-mass reference) and associated dates. These data are part of the N0c data package delivered by the CECT to the CMSM. N0c data contains the data of only one session and thus covers durations from a few orbits to 120 orbits. It is the starting point of the science flowchart. The orientation of the accelerometer with respect to the satellite has been precisely measured on ground and is sufficiently stable to be used all along the mission and set to a constant value in the data product. 

Three levels of data are used: N0 (raw data), N1 (pre-calibrated data) and N2 (science data used for calibration and for EP test). They are described in Section \ref{ssect_datalevels}. 

The raw data was provided by CECT to CMSM in ASCII files for each SU. They contained all test-mass acceleration measurements (at 4~Hz sampling rate) and housekeeping (at 1~Hz sampling rate). Other information related to the satellite status or its performance is also provided to CMSM.

A pre-processing at CMSM consisted in converting the data into binary files and in automatically creating associated masks at the N0 level (as shown below, masks evolve from the N0 to the N2 levels). Those masks characterise the validity of each measurements sample, with values set to 0 for data points to be ignored and 1 for valid data points.


The data processing is based on the measurement equation (which relates the measured acceleration signals to the Earth gravity signal and to the various instrumental parameters \cite{rodriguescqg1}). For more details, see Refs. \cite{rodriguescqg1, bergecqg7, metriscqg9}. In particular, Ref. \cite{bergecqg7} gives the performance obtained with mock data. 

\subsection {Singular events and missing data} \label{sing}

\begin{figure}
\includegraphics[width=0.9 \textwidth]{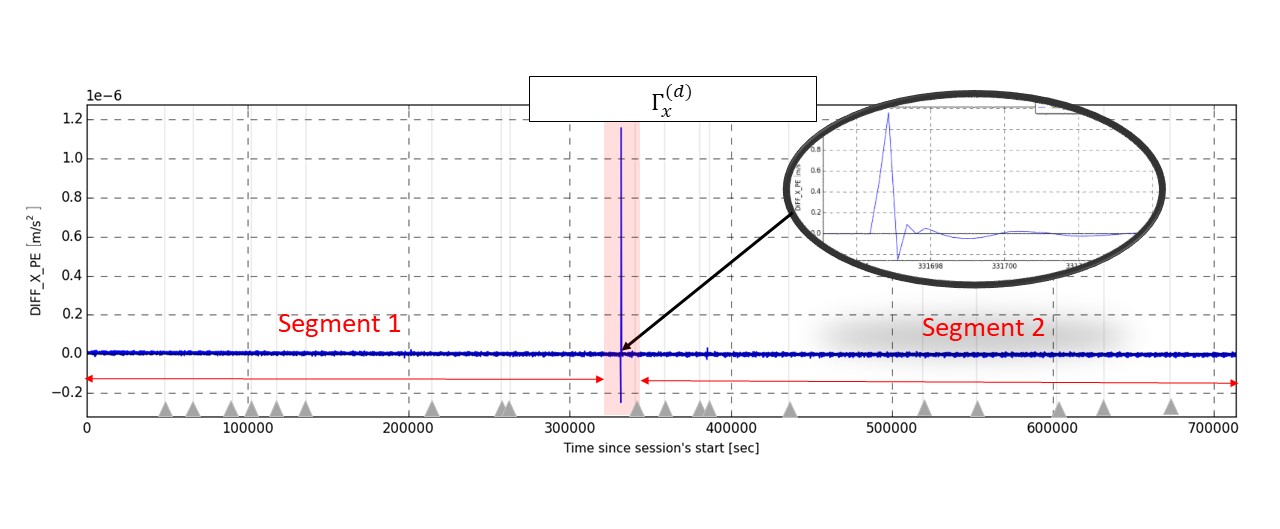}
\centering
\caption{Acceleration differential measurement with one major micro-debris peak, the position of which defines two segments. Grey triangles and lines show the position of each telemetry loss.}
\label{meteo}
\end{figure}

Data can be missing or discarded in the acceleration time series for different reasons:
\begin {itemize}
\item telemetry loss, though this is very rare (a few events per session, each with only one or two missing points, shown by grey triangles in \Fref{meteo});
\item the data does not meet some validity criterion (out of a predefined range, saturated data, detection of errors in the ICU, disturbing signal).
\end {itemize}

\begin{figure}
\includegraphics[width=0.5 \textwidth]{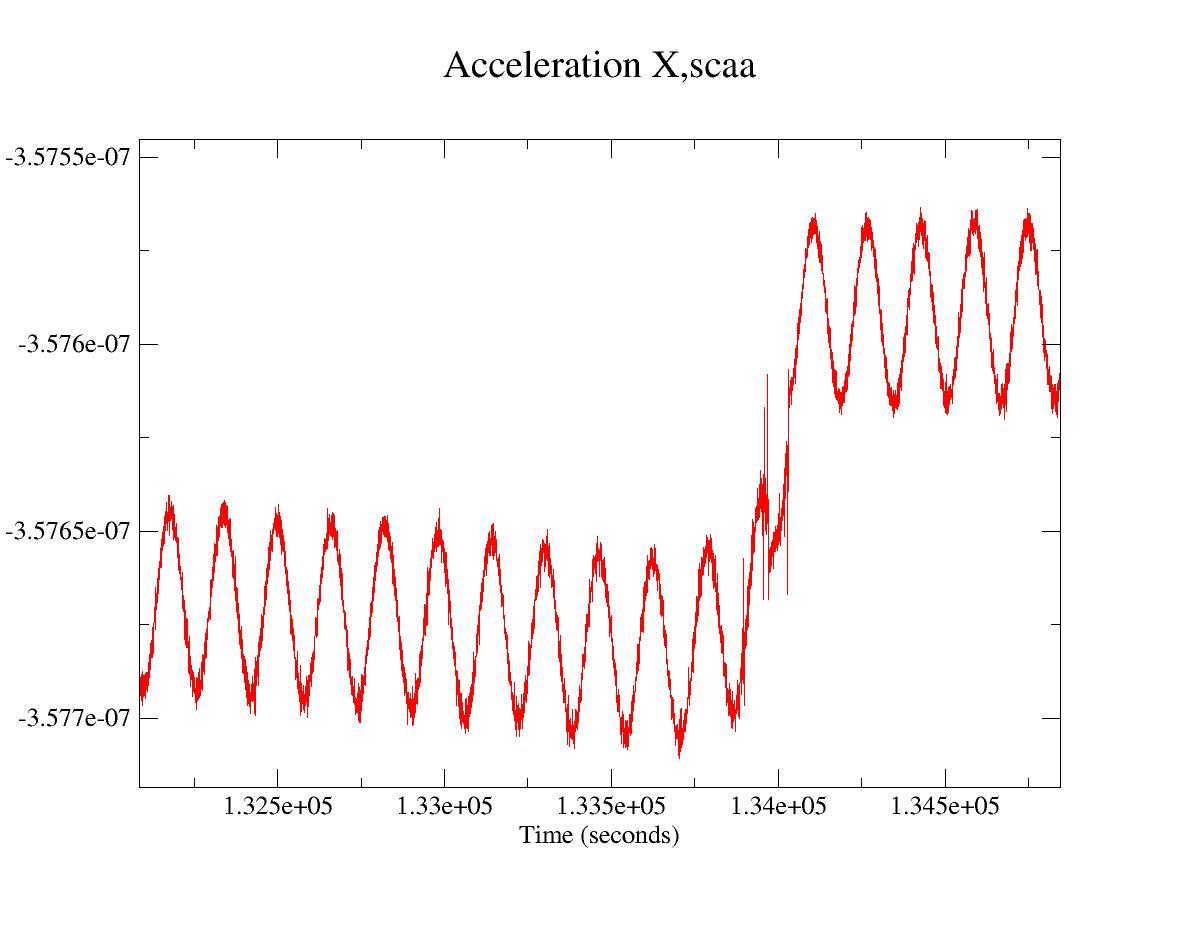}
\includegraphics[width=0.4 \textwidth]{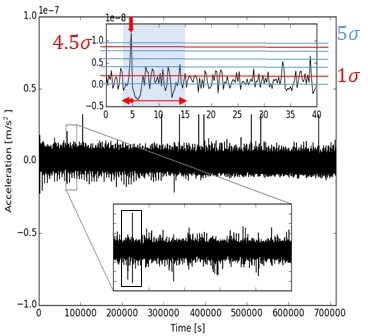}
\centering
\caption{Left: acceleration differential measurement of SUREF filtered by a low-pass filter. Right: typical glitch detected with a 4.5$\sigma$ clipping leading to suppress 16~s of data (in gray blue). }
\label{fig380}
\end{figure}

Micro-debris can generate large peaks in the acceleration measurement as illustrated in the middle of \Fref{meteo}. Nevertheless, such events are extremely rare.
Additionally, a few sudden changes in the local mean measured acceleration were observed in SUREF data (left panel of \Fref{fig380}). These leaps are not well understood as they are unpredictable, rare and correlated to no other observable events. We deal with these two types of event by discarding data around them. However, as the resulting gaps are wide (a few tens of seconds), we decided to split the concerned sessions in segments (see \Fref{meteo} for a session split into two segments). Segments' duration must follow the same rules as the session: as long as possible and including an even number of orbital periods. 

The most frequent events are ``glitches": small and transient variations of the measured accelerations (right panel of \Fref{fig380}). Similar events were observed in the GRACE mission and were attributed to thermomechanical spurious forces on the satellite structure \cite{flury08}. Few glitches were also observed at the sub pico-Newton level in LISA Pathfinder \cite{armano18}; their origin is still uncertain and could come from outgassing bursts. In MICROSCOPE, glitches are pretty large (1-10\,nm\,s$^{-2}$) and appear simultaneously on all test-masses.  They are most certainly due to crackles of the MLI (Multi-Layer Insulation) covering the satellite. Ref. \cite{bergecqg8} discusses their effect on the EP test and shows that they preferentially occur with a frequency equal to the EP test frequency $f_{\rm EP}$. However, glitches are not significantly subtracted in the differential measurements even after matching of the scale factors. Several hypotheses to explain this unexpected behaviour have been put forth and investigated but none allowed us to reach a firm conclusion: mismatching of transfer functions, non-linearities, saturation of signals in the accelerometer servo-loops, asymmetric gains of electronics. Investigations on some other potential sources of defects will be performed in the near future. 

As glitches have a signature at $f_{\rm EP}$, they compete with a potential EP violation signal. We defined the following algorithm to look for and mask glitches and remove most of their effect:
\begin{enumerate}
\item correct the differential acceleration with the calibrated parameters;
\item look for outliers along the $X$, $Y$ and $Z$ axes with a sigma-clipping technique on the calibrated differential acceleration, using a $4.5\sigma$ detection threshold;
\item flag data 1~s before and 15~s after each outlier (glitches emerge quickly from the noise then die off within a few seconds); 
\item repeat steps (ii) and (iii) on the calibrated differential acceleration filtered by a low pass Butterworth filter with a cut-off frequency of 0.01~Hz, with a $3\sigma$ detection threshold;
\item create a mask set to 0 on flagged data and 1 elsewhere and save it as a binary file;
\item estimate the E\"otv\"os parameter with MECM \cite{baghi16, metriscqg9} for each segment or session.
\end{enumerate}
The MECM process is detailed in Ref. \cite{baghi16, baghi16b}. The philosophy of the analysis method relies on an estimation of the noise in order to build an approximate generalized least square estimator that minimizes the variance. The key point is to get the estimation of the spectral distribution noise as close as possible to as the distribution without noise in order to minimize  leakage phenomenon. 
This process was tested by introducing fake EP violation signals in real data. The main concern was to evaluate the impact of removing data. Ref. \cite{metriscqg9} shows that fake EP signals (even of 10$^{-15}$) are only slightly impacted by this process: the error of estimation of the fake signal with MECM is less than 0.4\%.

\subsection{Data levels} \label{ssect_datalevels}

\subsubsection {N0 level}

N0 data are converted into binary format by CMSM from ASCII files delivered by CECT. The attitude and position of the satellite are kept in their ASCII origin format. At N0 level, the accelerometer data are expressed in each test-mass sensor frame. The ground segment accelerometer data contains the information used on-board as input by the DFACS as well as additional data headers, dates, electronics status and housekeepings.

This level comprises three sublevels: N0a, N0b and N0c. 

N0a is a short span data provided just after the satellite has passed over one of the ground stations. It was used only during critical operations (e.g. T-SAGE first switch on).  

N0b data are daily data spanning 24~h from UTC 00:00:00 and to UTC 23:59:59.99. They were available the day following the data acquisition. 

N0c data is the same as N0b but sewed and sorted by session. N0c data were delivered one or two weeks after the end of the associated session to allow for the completion of additional accurate information about the orbit and the attitude of the satellite estimated a posteriori. It includes complementary data of the DFACS performance for the related session as estimated by CECT \cite{robertcqg3}: angular and linear acceleration residuals at $f_{EP}$, 2$f_{EP}$ and at calibration frequencies.  

For the technical sessions, N0b\_S data level was produced in the same way as N0c data, but with no a posteriori improved orbit and attitude data. 

In addition to data used for the EP test, on-board measurements were available in order to characterise the environment (propulsion, temperature) or to monitor the good behaviour of the satellite. \Tref{N0_data} lists the type of data present in N0 data.

\begin{table}
\caption{\label{N0_data} List of data contained in the N0 data level.}
\centering
\begin{tabular}{@{}ll}
\br
Type of data  & Role  \\ 
\br
Executed scenario for the session & extract for data process  \\ \hline
Precise attitude of the satellite & science \\ \hline
Satellite and DFACS mode & science \\ \hline
8 thruster commands & DFACS, monitoring \\ \hline
Linear force DFACS commands about 3 axes & monitoring \\ \hline
Star-tracker quaternion & DFACS \\ \hline
13 Satellite temperature probes & characterisation  \\ \hline
3-axis linear accelerations of 4 test-masses &  DFACS, science \\ \hline
Precise acceleration along X of 4 test-masses & not usable for the EP test \\ \hline
3-axis angular accelerations of 4 test-masses &  DFACS, science \\ \hline
Linear and angular accelerometers flags & monitoring \\ \hline
Position of 4 test-masses along 3 axes & characterisation\\ \hline
Attitude of 4 test-masses about 3 axes & characterisation \\ \hline
6 temperatures in each SU & characterisation \\ \hline
5 temperatures in each FEEU & characterisation \\ \hline
3 temperatures in each ICU & monitoring \\ \hline
Reference voltage housekeeping & monitoring \\ \hline
Supply voltage housekeeping & monitoring \\ \hline
Accelerometer electronics commands and counters  & housekeeping  \\ \hline
Accelerometer electronics status & housekeeping  \\ \hline
Accelerometer software status and errors & housekeeping  \\ \hline
Satellite subsystem and accelerometer datings & 0.01\,Hz to 4\,Hz sampling \\ \hline

\end{tabular}
\end{table}

\subsubsection {N1 level}

The first step of the data processing is to produce N1a data from N0c data. It consists in changing the orientation of the accelerometer data measurement to a unique instrument reference frame defined as SUREF axes. SUEP and SUREF are mechanically identical but because of harness integration constraints, SUEP is turned by 180 degrees about the $X$ direction. As a consequence, SUEP measurements along $Y$, $Z$ and the rotations about these axes are inverted with respect to the SUREF ones at N0c level and must be turned by 180 degrees at N1a level.  
N1 level is mainly used for monitoring and does not contain any other sublevel except N1a for science sessions. For the technical sessions, N1a\_S data is produced similarly to N1a from N0b\_S instead of N0c.

\subsubsection{N2 level} \label{N2lev}

The N2 level is dedicated to science. It is subdivided in two sublevels: N2a and N2b. For technical sessions, only N2a\_S are produced from N1a\_S in the same way as N2a.

N2a is produced from N1a and combines individual test-mass accelerations in half difference accelerations $(\vv{\Gamma_1}-\vv{\Gamma_2})/2$ and mean accelerations $(\vv{\Gamma_1}+\vv{\Gamma_2})/2$ of the two test-masses (1) and (2), (1) being the inner test-mass and (2) the outer one of each SU. In Refs. \cite{rodriguescqg1, liorzoucqg2, robertcqg3, chhuncqg5, hardycqg6, bergecqg7, bergecqg8, metriscqg9}, ``differential acceleration'' is used for the difference of measurement acceleration $\vv{\Gamma^{(d)}} = (\vv{\Gamma_1}-\vv{\Gamma_2})$. N2a data include the Earth's gravity field $\vv g$ and gradient matrix $[T]$ which are computed from the ITSG-Grace2014s gravity potential model \cite{mayer06} expanded up to spherical harmonic degree and order 50. This computation is performed by the ACTENS software \cite{baghi16b}. The inertia gradient matrix ${\rm [In]}$ is computed at the same time. $\vv g$, $[T]$ and ${\rm [In]}$ are estimated by taking into account the satellite position, attitude and angular motion. 
The projection of the differential acceleration vector on the $X$ measurement axis can be approximated by \cite{rodriguescqg1} :
\begin{equation}  \label{eq_xacc}
\Gamma_x^{(d)} \approx  b_{x} + \tilde{a}_{cxx} \delta g_x + \sum_{i} \sum_{j} \left(T_{ij} - {\rm In}_{ij} \right) \tilde{a}_{cij} \Delta_j + 2\sum_{j} a_{dxj} \Gamma_j^{(c)} + n, \\
\end{equation}
where $i,j \in \{x,y,z\}$ and
\begin{itemize} 
\item $b_{x}$ is the total measurement bias projected on $X$; it can drift or fluctuate (systematic disturbing term) during the measurements;
\item $\delta$ is the E\"otv\"os parameter;
\item $\delta \vv{g}$ is projected on the $X$ measurement axis through the common-mode sensitivity matrix elements $\tilde{a}_{cij}$ ($\tilde{a}_{cjj} \approx 1$ and $\tilde{a}_{cij\ne i} \approx 0$);  
\item $\Delta_j$ is the ``offcentring" along the $j$ axis between the test-masses of the considered differential accelerometer;
\item $a_{dxj}$ is the first row component of the differential-mode sensitivity matrix that projects the mean non gravitationnal acceleration applied to the satellite (well approximated by the common mode acceleration $\Gamma^{(c)}$) on the $X$ axis;
\item $n$ represents all the stochastic disturbing terms \cite{rodriguescqg1, hardycqg6}.  
\end{itemize}

To estimate the E\"otv\"os ratio, we first estimate some of \Eref{eq_xacc}'s perturbing terms before correcting the measurement for them, to finally extract the signal in phase with the gravity field. These perturbing terms are systematic errors linked to the relative offcentring of the test-masses and to the differential mode sensitive matrix \cite{hardycqg6}.

Sessions are grouped in phases as already mentioned in \Tref{tab_scii} where the $a_{dxj}$ and $\Delta_y$ parameters are determined respectively with the calibration sessions ${\rm CAL}_{\rm{K1_{dx}}}$, ${\rm CAL}_{\rm{teta_{dZ}}}$, ${\rm CAL}_{\rm{teta_{dY}}}$ and ${\rm CAL}_{\rm{delta_Y}}$. The calibration method is detailed in Ref.\cite{hardycqg6}. It consists of applying a reference signal at $f_{\rm cal}$ in the DFACS loop for all ``CAL'' type sessions. This reference signal amplifies the contribution of one or the other parameter to be estimated. For instance, if the two test-mass outputs are biased by a sine signal along $X$, and thus generate a common mode acceleration at the DFACS input, the propulsion system applies the opposite acceleration , $\Gamma_x^{(c)}$, in order to cancel the accelerometer mean outputs. Differential acceleration is measured and the term $a_{dxx} \Gamma_x^{(c)}$ in \Eref{eq_xacc} is emphasized. 

From N2a data of calibration sessions, gaps coming from telemetry losses are replaced by the local mean. It must be noted that all calibration sessions selected to calibrate EP sessions are cleaned from major events like micro-debris impulses or leaps, described in \Sref{sing}. For the few orbits of the calibration sessions, the small drift is fit by a first-order polynomial and subtracted before running ADAM, a least square regression in the frequency domain well detailed in Ref. \cite{bergecqg7, metriscqg9}. The four parameters are estimated simultaneously in an iterative process with ADAM: the data is corrected with the estimated parameter of the previous iteration for a better estimation. Five iterations are used to converge on a precise estimation \cite{hardy13c}. For each phase a pre-calibration matrix of parameters is produced. 

The next step consists in estimating $\Delta_x$ and $\Delta_z$ to complete the pre-calibration matrix. From N2a data of EP sessions, minor peaks and data holes are detected and replaced by the local mean. If a major peak or leap is present, the session is split in two segments (\Sref{sing}). The data is corrected by the pre-calibration matrix and a third-order polynomial is fit to correct the trend in the time series.  $\Delta_x$ and $\Delta_z$ are then estimated for each segment with ADAM. Hence, the pre-calibration matrix can be completed in a calibration matrix associated to each segment. \Tref{tab-segments} gives the list of the selected sessions with the attribute ``-1" and ``-2" when the session is split into two segments. In this table, the SUEP sessions \#148 and \#160 are not listed because they do not fulfil the criteria of a non-linearity coefficient lower than $14000$\,s$^{-2}$ \cite{hardycqg6, metriscqg9}. SUEP session \#430 was also discarded because it was performed $20^\degree$C higher than the other sessions: this session was used to check the temperature sensitivity for test-mass relative off-center positions. Session \#452 and \#454 are also excluded for SUEP but not for SUREF because the DFACS operates with SUREF in these sessions leading to a less optimal acceleration environment for SUEP.

Once the parameters are calibrated, the EP measurement data at N2a level are corrected for the effect of $\Gamma_j^{(c)}$ and for the gravity gradient effects with the calibrated parameters. This step produces calibrated data for each session or for the two segments of the session when necessary. The sigma clipping detection of glitches detailed in \Sref{sing} is applied to these data with eliminated data identified by a mask. The calibrated and cleaned data is computed by MECM to extract the E\"otv\"os parameter and to produce calibrated filled data: missing data are replaced by MECM with data based on the best estimate of the noise \cite{metriscqg9}. 

Finally these calibrated and filled data are processed by ADAM in a single run over all pre-processed segments. MECM has been shown to be robust to missing data and efficient to fill gaps in order to enable ADAM to estimate the E\"otv\"os parameter with all sessions in a single run.  Ref. \cite{metriscqg9} shows that ADAM and MECM provide consistent estimations of the E\"otv\"os parameter, though they are fundamentally different algorithms. Theoretically, MECM could be also applied on all data at once but it is much more expensive in time and memory. As an example, on only 120 orbits-data, it took 12 to 24 hours with 40 Gbytes of memory for MECM to converge versus 10 min and 1 GByte for ADAM. The expense in memory varies with the square of number of orbits for MECM. Using MECM to compute more than 1000 orbits would require much more powerful computers for a comparable result to ADAM as discussed in Ref.\cite{metriscqg9}.

\begin{table} [H]
\caption{\label{tab-segments}Characteristics of the segments selected for the EP sessions. The segment number corresponds to the session number extended by an index related to segment number in the session. The duration is given as a multiple of orbital periods, keeping in mind that this period is about 5946 s. The third column indicates the percentage of data eliminated during the pre-processing (see \Sref{N2lev}).} 
\begin{indented}
\lineup
\item[]\begin{tabular}{@{}*{3}{lll}}
\br                              
$\0\0$Segment &Duration &Percentage of data \cr 
$\0\0$number  &(orbits)    &  eliminated (glitches) \cr 
\br
\multicolumn{3}{c}{SUREF} \cr
\br
\0\0120-1&22 & 4   \cr
\0\0120-2&64 & 15 \cr
\0\0174    &86 & 25 \cr
\0\0176    &62 & 40 \cr
\0\0294    &76 & 17 \cr
\0\0376-1&36 & 14 \cr
\0\0376-2&28 & 11 \cr
\0\0380-1&46 & 7   \cr
\0\0380-2&34 & 5   \cr
\0\0452    &32 & 20 \cr
\0\0454    &56 & 22 \cr
\0\0778-1&38 &0   \cr
\0\0778-2&18 & 6  \cr 
\br
\multicolumn{3}{c}{SUEP} \cr
\br                              
\0\0210    &50   &18 \cr
\0\0212    &60   & 17 \cr
\0\0218    &120 & 15 \cr
\0\0234    &92   & 18 \cr
\0\0236    &120 & 21 \cr
\0\0238    &120 &24 \cr
\0\0252    &106 & 26 \cr
\0\0254    &120 & 27 \cr
\0\0256    &120 & 28 \cr
\0\0326-1&66   & 12 \cr
\0\0326-2&34   & 7  \cr
\0\0358    &92   & 14 \cr
\0\0402    &18   & 35 \cr
\0\0404    &120 & 23 \cr
\0\0406    &20   & 23 \cr
\0\0438    &32   & 21 \cr
\0\0442    &40   & 21 \cr
\0\0748    &24   & 25 \cr
\0\0750    &8     & 19 \cr
\br
\end{tabular}
\end{indented}
\end{table}



\section{Conclusion} \label{sect_ccl}

The mission has been managed in cooperation between CNES, OCA and ONERA. The Science Performance Group including also the ZARM institute had gathered regularly for more than 15 years in order to monitor all the science analysis. This activity contributed to define and validate all the processes needed to prepare the science data. The Science Working Group composed of experts not directly involved in the definition of the mission met regularly during the mission definition and after the launch for the mission scenario strategy. The a priori mission scenario had to be updated to take into account new constraints imposed by unexpected events and additional experiments in orbit. In particular a capacitor failure in the SUREF accelerometer led to minimise the operating time of each instrument and to reduce the operating temperature for a lifetime optimisation. 

Calibration sessions were realised throughout the mission to evaluate any evolution in the instrument parameters and to provide calibrated differential measurement data. In 2018, long thermal tests were performed, over more than 500 orbits, to improve the thermal model of the experiment and to better estimate the related systematic errors \cite{hardycqg6}.  
The data science process was tested before the launch and updated during the mission to validate the robustness of the analysis and cope with unexpected events such as glitches. 
The presence of glitches led us to eliminate some degraded data up to 40\% for some sessions and to use an optimised process to cope with missing data \cite{baghi16b}. Finally, more than 1500 orbits worth of data were available to improve the laboratory EP tests \cite{schlamminger08, wagner12, williams12,viswanathan18} and those from the first paper in Ref.\cite{touboul17} based only on 120 orbits. Two softwares, ADAM and MECM, were used in Ref. \cite{metriscqg9} to estimate the E\"otv\"os parameter and gave consistent results. 
 
The mission data will be distributed from a dedicated server and made available during 5 years. All the data and additional information leading to the final result will be also made available to the scientific community.

\clearpage

\ack

The authors express their gratitude to all the different services involved in the mission partners and in particular CNES, the French space agency in charge of the satellite. This work is based on observations made with the T-SAGE instrument, installed on the CNES-ESA-ONERA-CNRS-OCA-DLR-ZARM MICROSCOPE mission. ONERA authors’ work is financially supported by CNES and ONERA fundings.
Authors from OCA, Observatoire de la C\^ote d’Azur, have been supported by OCA, CNRS, the French National Centre for Scientific Research, and CNES. ZARM authors’ work is supported by the German Space Agency of DLR with funds of the BMWi (FKZ 50 OY 1305) and by the Deutsche Forschungsgemeinschaft DFG (LA 905/12-1). The authors would like to thank the Physikalisch-Technische Bundesanstalt institute in Braunschweig, Germany, for their contribution to the development of the test-masses with funds of CNES and DLR. 


\appendix
\section {Detailed realised mission scenario} \label{sect_tabscen}

In this appendix, we list all MICROSCOPE's operations. Sessions are gathered in calendar years.

\begin{table} 
\caption{\label{tab-sessions16}List of sessions with T-SAGE ON performed in 2016: odd session numbers are transition sessions and are not reported here. $\P{}$: first and second capacitance breakdown. $\dag$: Session interrupted by a DFACS equipment anomaly.} 

\begin{indented}
\lineup
\item[]\begin{tabular}{@{}rllcc}
\br                              
Start & Type & Session& Duration&SU        \cr 
orbit &         & numbers & (orbits)&on    \cr 
\mr
91&Commissioning&4 to 20&217&both        \cr
399&Commissioning&24&14&both             \cr
413&Commissioning&26 to 52&334&both       \cr
748&$\rm{EP_I}$&54$^{\P{}\dag}$&17&SUREF      \cr
1820&Commissioning&62-64&17&SUREF       \cr
1838&$\rm{EP_I}$&66$^{\P{}\dag}$&7&SUREF      \cr
1846&Standby&68&13&SUREF       \cr
1859&CAL&70&5&SUREF       \cr
1866&$\rm{EPR_{V1}}$&72&20&SUREF       \cr
2138&Commissioning&76&19&SUEP       \cr
2158&CAL&78&5&SUEP       \cr
2165&$\rm{EPR_{V2}}$&80$^{\dag}$&18&SUEP      \cr
2246&CAL&84&5&SUEP       \cr
2253&$\rm{EPR_{V2}}$&86&120&SUEP      \cr
2375&CAL&88 to 106&80&SUEP       \cr
2465&Commissioning&108&28&SUEP       \cr
2651&Commissioning&112&15&SUREF       \cr
2667&CAL&114&5&SUREF      \cr
2674&$\rm{EPR_{V2}}$&116$^{\dag}$&2&SUREF       \cr
2681&$\rm{EPR_{V2}}$&120&120&SUREF     \cr
2804&CAL&122 to 128$^{\dag}$&20&SUREF      \cr
2830&Stanby&132&26&SUREF       \cr
3076&CAL&146&5&SUEP       \cr
3084&$\rm{EPR_{V3}}$&148$^{\dag}$&23&SUEP    \cr
3107&Stanby&150 to 156&98&SUEP       \cr
3207&CAL&158&5&SUEP       \cr
3215&$\rm{EPR_{V3}}$&160&120&SUEP       \cr
3338&CAL&162 to 164&10&SUEP       \cr
3402&PID tests&168&18&SUEP       \cr
\br
\end{tabular}
\end{indented}
\end{table}

\begin{table}
\caption{\label{tab-sessions17}Same as \ref{tab-sessions16}, for sessions performed in 2017. Sensitivity stands for thermal sensitivity sessions. Characterisation stands for technical sessions testing the instrument as the PID, the loop transfer function, the stiffness or the test-mass free-motion.}
\begin{indented}
\lineup
\item[]\begin{tabular}{@{}rllcc}
\br                              
Start & Type & Session& Duration&SU         \cr 
orbit &         & numbers & (orbits)&on     \cr 
\mr
3890&CAL&172&5&SUREF       \cr
3897&$\rm{EPR_{V2}}$&174 to 176&202&SUREF      \cr
4104&CAL&178 to 192&44&SUREF       \cr
4155&PID test&194&18&SUREF       \cr
4174&CAL&196 to 202&4&SUREF      \cr
4267&CAL&206&10&SUEP       \cr
4281&$\rm{EPR_{V3}}$&210 to 212$^\dag$&126&SUEP     \cr
4472&CAL&216&5&SUEP        \cr
4480&$\rm{EPR_{V3}}$&218&120&SUEP       \cr
4602&CAL&220 to 226 &20&SUEP      \cr
4688&CAL&232 &5&SUEP       \cr
4693&$\rm{EPR_{V3}}$&234 to 238&332&SUEP       \cr
5033&CAL&240 to 246 &20&SUEP      \cr
5113&CAL&250 &5&SUEP       \cr
5121&$\rm{EPR_{V3}}$&252 to 256&346&SUEP       \cr
5473&CAL&258 to 264&20&SUEP      \cr
5497&Sensitivity&266 to 270 & 34 & SUEP  \cr
7211&PID test&282&6&SUREF    \cr
7316&CAL&286 to 292&20&SUREF       \cr
7342&$\rm{EPR_{V3}}$&294&94&SUREF       \cr
7439&CAL&296&5&SUREF      \cr
7444&Sensitivity&298 to 300 & 5 & SUREF \cr
7450&CAL&302&5&SUREF       \cr
7455&Sensitivity&304 to 306 & 8 & SUREF \cr
7465&CAL&308&5&SUREF      \cr
7511&CAL&312&5&SUEP       \cr
7516&Sensitivity&314 to 316 & 5 & SUEP \cr
7522&CAL&318&5&SUEP       \cr
7527&Sensitivity&320 to 322 & 8 & SUEP \cr
7537&CAL&324&5&SUEP       \cr
7545&$\rm{EPR_{V3}}$&326&102&SUEP      \cr
7702&CAL&330 to 340$^{\dag}$ &25&SUEP      \cr
7757&CAL&344 to 356 &30&SUEP       \cr
7802&$\rm{EPR_{V3}}$&358$^{\dag}$&92&SUEP       \cr
7895&Characterisation&360 to 370 &3&SUEP       \cr
8195&CAL&374&5&SUREF      \cr
8203&$\rm{EPR_{V2}}$&376&80&SUREF       \cr
8284&CAL&378&40&SUREF      \cr
8327&$\rm{EPR_{V3}}$&380&120&SUREF       \cr
8450&CAL&382 to 396&40&SUREF       \cr
8551&CAL&400&5&SUEP       \cr
8559&$\rm{EPR_{V2}}$&402&20&SUEP       \cr
8582&$\rm{EPR_{V3}}$&404 to 406&140&SUEP       \cr
8728&CAL&410 to 414&15&SUEP       \cr
\br
\end{tabular}
\end{indented}
\end{table}

\begin{table}
\caption{\label{tab-sessions18}Same as \ref{tab-sessions16}, for sessions performed in 2018. Sensitivity stands for thermal sensitivity sessions. Characterisation stands for technical sessions testing the instrument as the PID, the loop transfer function, the stiffness or the test-mass free-motion. $\P{}$: third capacitance breakdown.}
\begin{indented}
\lineup
\item[]\begin{tabular}{@{}rllcc}
\br                              
Start & Type & Session& Duration&SU        \cr 
orbit &         & numbers & (orbits)&on     \cr 
\mr
8977&Characterisation&416 to 420&3&SUEP with SUREF on       \cr
8981&CAL&422 to 428&20&SUEP with SUREF on       \cr
9007&$\rm{EPR_{V2}}$&430$^{\dag}$&17&SUEP with SUREF on     \cr
9152&CAL&436&5&SUEP       \cr
9160&$\rm{EPR_{V2}}$&438&40&SUEP    \cr
9201&CAL&440&40&SUEP       \cr
9243&$\rm{EPR_{V2}}$&442&40&SUEP    \cr
9286&CAL&444&5&SUEP      \cr
9261&Sensitivity&446 to 448&152&both       \cr
9444&CAL&450&5&SUREF with SUEP on       \cr
9452&$\rm{EPR_{V2}}$&452 to 454$^{\P{}\dag}$&100&SUREF with SUEP on      \cr
9552&Standby&456 to 458&6&SUEP       \cr
9932&Characterisation&474 to 520&31&SUEP       \cr
9264&CAL&522&5&SUEP       \cr
9969&Characterisation&524 to 562&46&SUEP       \cr
10015&Sensitivity&564 to 566&72&SUEP       \cr
10090&Standby&end of 566&1&SUREF\cr
10090&Sensitivity&568&20&SUREF       \cr
10110&Characterisation&570 to 650&72&SUREF       \cr
10182&Standby&start of 652&1&SUREF\cr
10183&Standby&652&90&SUEP\cr
10273&Characterisation&654 to 660&40&SUEP\cr
10313&Standby&662&48&SUEP\cr
10361&Characterisation&664&20&SUEP\cr
10417&Sensitivity&666 to 684&772&SUEP\cr
11189&Standby&686 to 704&396&SUEP\cr
11585&Sensitivity&706&101&SUEP\cr
11686&Standby&706 to 738&792&SUEP\cr
12479& CAL & 740 to 746 & 20 & SUEP  \cr
12506&$\rm{EPR_{V2}}$&748&24&SUEP    \cr
12533&$\rm{EPR_{V3}}$&750&8&SUEP    \cr
12543&Characterisation&752 to 760&20&SUEP       \cr
12853&CAL&772 to 776&15&SUREF  \cr
12874&$\rm{EPR_{V2}}$&778&60&SUREF     \cr
12936&Characterisation&780 to 786 &6&SUREF       \cr
13068& End of mission &798 & 18& SUEP \cr
\br
\end{tabular}
\end{indented}
\end{table}

\clearpage
\section*{References}
\bibliographystyle{iopart-num}
\bibliography{biblimscope}

\end{document}